\documentclass[aps,prb,reprint,nofootinbib,twocolumn,superscriptaddress,showpacs,showkeys,longbibliography,amsmath,amssymb]{revtex4-2}
\usepackage{graphicx}
\usepackage{dcolumn}
\usepackage{bm}
\usepackage{braket}
\usepackage{subfigure}
\usepackage[colorlinks,bookmarks=false,citecolor=blue,linkcolor=red,urlcolor=blue]{hyperref}
\usepackage[english]{babel}
\usepackage{amssymb}
\usepackage{epstopdf,epsfig}
\usepackage{lipsum}
\usepackage{times} 
\usepackage{helvet} 
\usepackage{courier} 

\makeatletter

\newcommand{\Rmnum}[1]{\expandafter\@slowromancap\romannumeral #1@}
\makeatother

\begin{document}
\preprint{APS/123-QED}

\title{Hybrid skin-topological effect in non-Hermitian checkerboard lattices with large Chern numbers} 

\author{Yi-Ling Zhang}
\affiliation{School of Physical Science and Technology, and Collaborative Innovation Center of Suzhou Nano Science and Technology, Soochow University, 1 Shizi Street, Suzhou 215006, China}
\author{Li-Wei Wang}
\affiliation{School of Physical Sciences, University of Science and Technology of China, Hefei, 230026, China}
\author{Yang Liu}
\affiliation{School of Physical Science and Technology, and Collaborative Innovation Center of Suzhou Nano Science and Technology, Soochow University, 1 Shizi Street, Suzhou 215006, China}
\author{Zhao-Xian Chen}
\affiliation{National Laboratory of Solid State Microstructures, Collaborative Innovation Center of Advanced Microstructures, and College of Engineering and Applied Sciences, Nanjing University, Nanjing 210093, China}
\author{Jian-Hua Jiang}
\email{jhjiang3@ustc.edu.cn}
\affiliation{School of Physical Science and Technology, and Collaborative Innovation Center of Suzhou Nano Science and Technology, Soochow University, 1 Shizi Street, Suzhou 215006, China}
\affiliation{School of Biomedical Engineering, Division of Life Sciences and Medicine, University of Science and Technology of China, Hefei 230026, China}
\affiliation{Suzhou Institute for Advanced Research, University of Science and Technology of China, Suzhou, 215123, China}

\date{\today}

\begin{abstract}
Non-Hermitian topology provides a research frontier for exploring topological phenomena, revealing novel topological effects and driving the development of emergent materials and platforms. Here, we explore the non-Hermitian Chern insulator phases and the hybrid skin-topological effects in checkerboard lattices with synthetic gauge fluxes. Such lattices can be realized in integrated silicon photonic nanocircuits and microresonators as well as in arrays of evanescently coupled helical optical waveguides. With a simple and tunable design, the system is found to support non-Hermitian hybrid skin topological effects, exhibiting corner skin effects when the lattice symmetry either $C_4$ or $C_2$. An unconventional physical mechanism is revealed as the origin of such a transition which is connected to the corner-induced scattering between the multiple chiral edge channels. These properties are enabled by the large Chern number and the rich non-Hermitian topological edge states in our system, revealing the diverse non-Hermitian topological bulk-boundary correspondence. Our design offers excellent controllability and experimental feasibility, making it appealing for studying non-Hermitian topological phenomena.
\end{abstract}

\maketitle

\section{Introduction}

Non-Hermitian physics provides an efficient framework for investigating non-equilibrium and open quantum systems as well as their classical counterparts, wherein dissipation, energy gain, and nonreciprocal couplings give rise to physical effects that are unattainable in conventional Hermitian systems. Non-Hermitian effects were found to result in versatile non-Hermitian phenomena such as the non-Hermitian skin effect (NHSE)~\cite{skin-effect1, skin-effect2, skin-effect3, skin-effect4, skin-effect5}, the exceptional points, lines and surfaces~\cite{non-Hermitian1, non-Hermitian2, non-Hermitian3, non-Hermitian4, non-Hermitian5,non-Hermitian6, non-Hermitian7, non-Hermitian8, non-Hermitian9}, the non-Hermitian topological band theory~\cite{non-hermitian-topology1, non-hermitian-topology2, non-hermitian-topology3,non-hermitian-topology4}, non- Bloch band theory~\cite{non-bloch1,non-bloch2,non-bloch3, non-bloch4}, unidirectional invisibility~\cite{feng2013experimental}, and non-Hermitian states permutation~\cite{10.1093/nsr/nwac010}. The NHSE can be characterized by the spectral topology in the complex plane for finite non-Hermitian systems\cite{PhysRevLett.124.086801,PhysRevLett.125.126402}. Non-Hermitian skin modes coexist with topological edge modes and thus extends the bulk-edge correspondence to a more general picture\cite{PhysRevLett.125.226402}. A particularly promising frontier lies at the interface between topological physics and non-Hermitian physics. The interplay between NHSE and topological edge modes can bring interesting physics, such as hybrid topological-skin effects~\cite{hybrid-skin-effect1, hybrid-skin-effect2, hybrid-skin-effect3, hybrid-skin-effect4}, where corner localization is due to the coaction of topological localization and NHSE.

Floquet non-Hermitian systems, with time-periodic modulation, provide a rich territory for exploring dynamic and topological phenomena~\cite{Floquet-non-Hermitian1,Floquet-non-Hermitian2, Floquet-non-Hermitian3, Floquet-non-Hermitian4, Floquet-non-Hermitian5}. Floquet systems can enable ``synthetic dimensions" and create a platform to engineer desired effective Hamiltonian with unconventional properties~\cite{Lustig2019}. Furthermore, Floquet systems can generate interesting spectral and dynamic properties, allowing for the emergence of various topological phases~\cite{Rechtsman2013, Wintersperger2020, PhysRevB.98.205417, PhysRevLett.123.190403}. The periodic modulation in Floquet systems introduces unique artificial gauge fields, enabling the manipulation of topological phases and the emergence of hybrid modes within a single setup. The hybrid topological-skin effects can be regarded as a second-order NHSE, where the topological edge states exhibit NHSE towards certain corner boundaries. However, the existing studies on hybrid topological-skin effects focus on the relatively simple regime where the spatial symmetry dictates the phenomena and there is only one branch of edge states. The hybrid topological-skin effects remain unknown in the regime with multiple branches of edge states where richer emergent phenomena are expected.

In this work, we study the interplay between the NHSE and the topological physics in Floquet non-Hermitian checkerboard lattices. At the heart of our theory is the non-Hermitian topological phases induced only by gain and loss as well as their interplay with the artificial gauge fields. For concreteness, we design a coupled photonic waveguide lattice to realize the theoretical model, where the synthetic gauge flux is generated through the Floquet modulation. Nevertheless, the underlying physics is more general and is not restricted to photonic systems. For instance, the same physics can be realized also in cold atom systems where the synthetic gauge flux can be realized via other mechanisms. We find that our checkerboard lattices can efficiently generate non-Hermitian bands with large Chern numbers. The induced multiple chiral edge states have rich non-Hermitian skin effects due to that each branch of the chiral edge states have its own non-Hermitian properties. The total hybrid topological-skin effect comes from the NHSEs of the multiple chiral edge states as well as their interaction. We find that the scattering between these multiple non-Hermitian topological edge channels at the corner boundaries can induce an unconventional mechanism for the corner skin effect, which is distinct from the previous mechanisms solely associated with the nonreciprocal properties of the non-Hermitian edge states. With a simple and tunable design, the system is found to support non-Hermitian hybrid skin topological effects, exhibiting rich corner skin effects when the lattice symmetry is either $C_4$ or $C_2$.

The structure of this paper is organized as follows: In Sec. \ref{The 2D photon-electron (SSH) coupling model}, we introduce the non-Hermitian effective model of photonic checkerboard lattices and the topological phases, which provide the groundwork for realizing the NHSE. In Sec. \ref{sec:c4 symmetry}, we investigate the mixed skin-topological effects in the non-Hermitian effective model with $C_4$ symmetry. We analyze the origin of corner skin modes from a dynamical perspective. The locations of different corner skin modes are influenced by various non-equilibrium interactions of topological edge states. In Sec. \ref{sec:c2 symmetry}, we explore the non-Hermitian effective model with $C_2$ symmetry. Corner skin modes still exist, and we present the dynamic analysis process, finding that the locations of corner skin modes depend not only on the signs of group velocity and the gain/loss of topological edge states, but also on the contributions from the topological edge states. Finally, we conclude in Section \ref{Conclusion}.
\section{NON-HERMITIAN PHOTON-ELECTRON COUPLING MODEL AND  CHERN INSULATOR PHASES}
\label{The 2D photon-electron (SSH) coupling model}
\begin{figure}    
\centering
\includegraphics[width=0.5\textwidth]{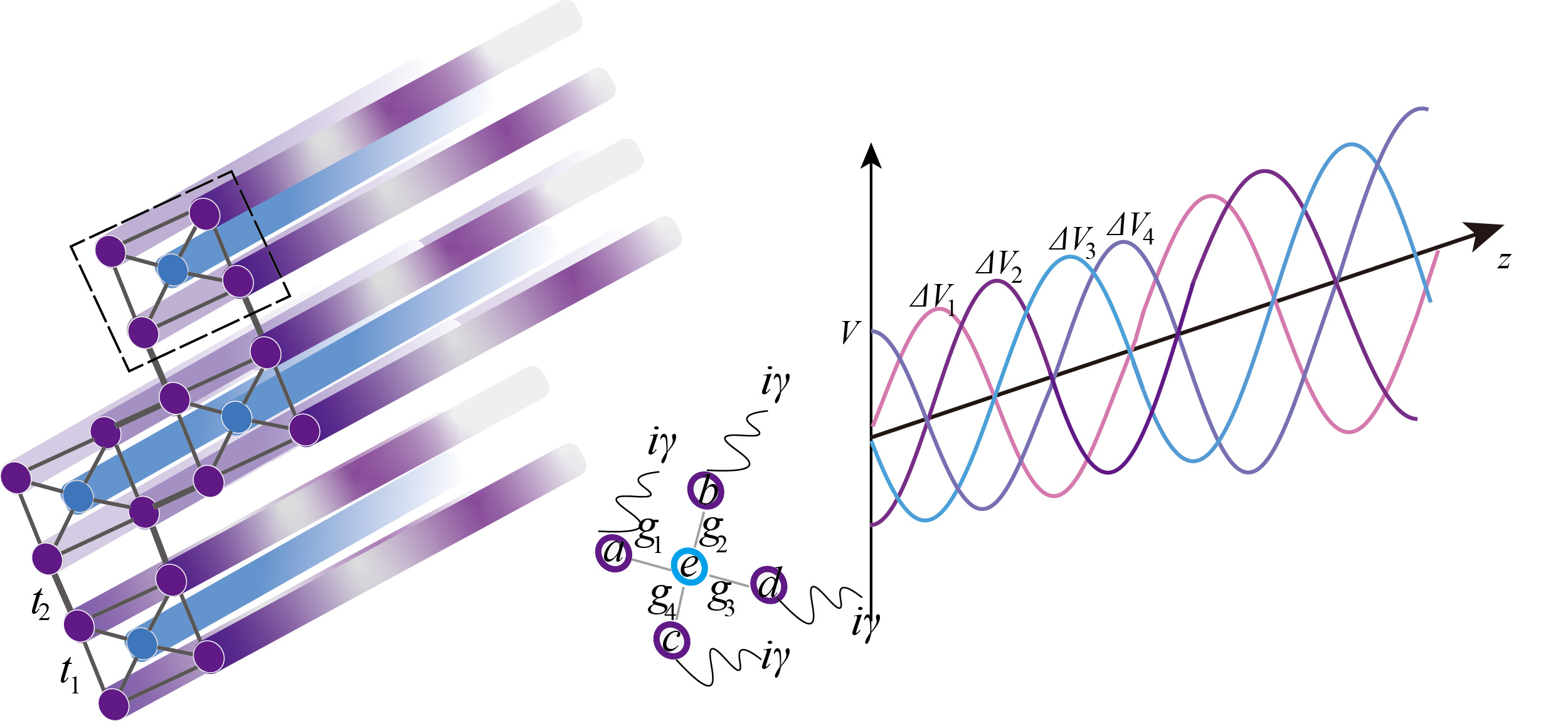}
\caption{ Floquet potentials in a chain square lattice (left panel). A unit-cell (maked by the dotted box) has five sites (labeled as a, b, c, d, e). The four sites, with lossy, for the on-site potential $\Delta V$ are modulated in sinusoidal form (right panel). Such  Floquet NHSE systems offer rich non-Hemition topological phase and states in a geometrically static arrangement.}
\label{fig:floquet}
\end{figure}
We introduce interstitial sites between the main sites of the two-dimensional square potential or Su-Schrieffer-Heeger (SSH) lattice , as shown in Fig. \ref{fig:floquet} . In each unit cell (marked by the dot-dashed box), an interstitial site (e) is added in the intermediate region between the four sublattice sites (a, b, c, d), forming a coupling path between each pair of sublattice sites. There is an inhomogeneous dissipation distribution and an imbalanced on-site potential in each unit cell. In this case, The $z$-dependent Hamiltonian for each of this model is partitioned as 
\begin{equation}
 {H}(z)={H}_{\text{SSH}}+H_{\text{loss-potential}}(z)+{H}_{\text{inter-SSH}}(z),
 \label{Hamiltonian_total}
\end{equation}
where ${H}_{\text{SSH}}$ and $H_{\text{loss-potential}}(z)$ respectively describe the main sites and their dissipation and  on-site potential distributions,  whereas ${H}_{\text{inter-SSH}}$ is about the coupling between the interstitial sites and main sites sublattice. We take 
\begin{align}
&H_{\text{SSH}}=\sum_{i,j}t_{1}(a_{i,j}^{\dagger}b_{i,j}+d_{i,j}^{\dagger}c_{i,j}+b_{i,j}^{\dagger}c_{i,j}+a_{i,j}^{\dagger}d_{i,j})\notag\\
&+\sum_{i,j}t_{2}(a_{i+1,j}^{\dagger}b_{i,j}+d_{i+1,j}^{\dagger}c_{i,j}+b_{i,j+1}^{\dagger}c_{i,j}+a_{i,j+1}^{\dagger}d_{i,j})\notag\\
&+\text{h.c.},
 \label{Hamiltonian_symmetry}
 \end{align}
where $t_1$ and $t_2$ represent the intra-cell and inter-cell coupling strengths in the 2D SSH lattice, respectively. $\xi_{i,j}^\dagger(\xi_{i,j})(\xi\in{\text{a, b, c, d}})$ creates (annihilates) operator and h.c. denotes the Hermitian conjugate. The coupling interaction between the two sub-systems, with $H_{\text{inter-SSH}}(z)=\sum_{i,j}\sum_{m=1}^{4}g_{m}(z)(e_{i,j}^{\dagger}\xi_{i,j})+\text{h.c.}$,
The clockwise rotation involves time-dependent nearest-neighbor interactions (with couplings $g_1(z), g_2(z), g_3(z), g_4(z)$), as shown in Fig.\ref{fig:floquet}. The on-site potential  at the sublattice sites (a, b, c, d), with dissipation $\gamma$, are given by 
\begin{align}
H_{\text{loss-potential}}(z)&=\sum_{i,j}(V_{i,j}(z)+i\gamma)\xi_{i,j}^\dagger\xi_{i,j}+\text{h.c.},
 \label{Hamiltonian_photon}
\end{align}
we consider a sinnusoidal modulation $V_{i,j}(z)=\text{sin}(2\pi z/T+2\pi m/4)$, with a chiral phase shift of $2\pi/4$ between the interstitial sites in the intermediate region of each main site, where $m\in{\text{1, 2, 3, 4}}$ and $T$ denotes the Floquet period along the $z$ direction. In the periodically Floquet systems, the static effective Hamiltonian $H_{\text{eff}}$ defined as
 \begin{align}
H_{\text{eff}}&=i/T\text{log}(U),
 \label{Hamiltonian_photon}
\end{align}
where $U=\mathcal{T}\text{exp}(-i\int_{0}^TH(z)dz)$ describes through the evolution operator $U$ for one period $T$, and $\mathcal{T}$ is the time-ordering. In the high driving frequency regime, $H_{\text{eff}}$  can be as a perturbation expansion in powers of $1/T$ (see details in the appendix \ref{sec: effective hamiltonian}).

\begin{figure}    
\centering
\includegraphics[width=0.5\textwidth]{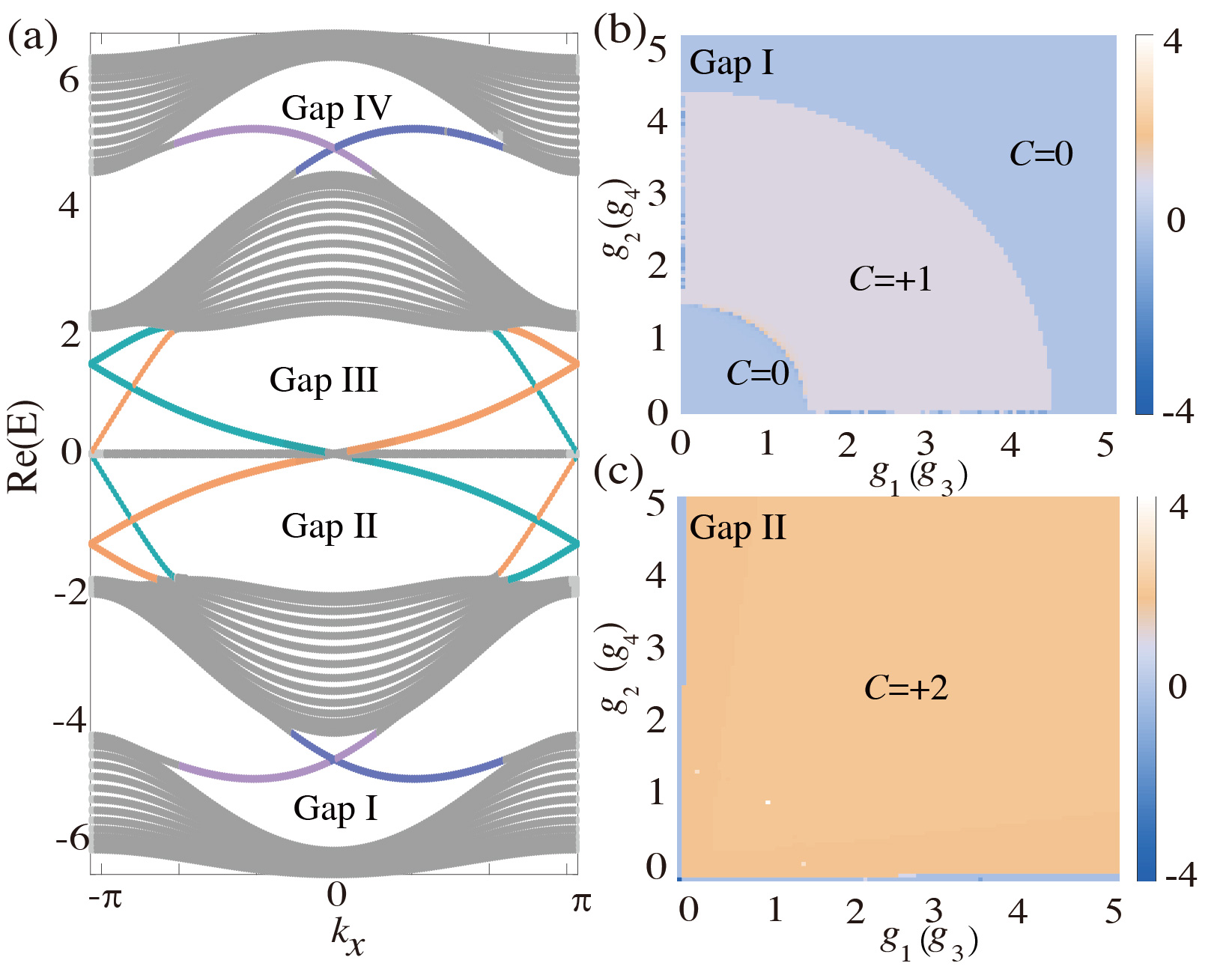}
\caption{The band structure and topological phase of the non-Hermitian effective model. (a) The real part of the projected band structure  calculated from the ribbon supercell with periodic boundary conditions in the $x$-direction and open boundary conditions in the $y$-direction. The parameters are set as $g_1=g_3=1, g_2=g_4=2$, $\gamma=-1$, and $\theta_{m}=(i-1)\pi/2$, where $m=1, 2, 3, 4$. (b-c) The real-space Chern numbers calculated for gaps I and gaps II as functions of \(g_1 (g_3)\) and \(g_2 (g_4)\) under loss \(\gamma = [0,1]\). The Chen numbers are distinguished by different colors.}
\label{fig:two_ssh_band1}
\end{figure}

 Next, we analyze whether the edge states are topological in effective model. Generally, topological invariants are used to characterize them. We calculate the real part of the projected band structure, as shown in Fig. \ref{fig:two_ssh_band1}(a). The band structure contains five bulk bands, with gaps I and IV (II and III) each containing one (two) pair of edge states. The edge states appearing in different band gaps are distinguished by different colors. Notably, unlike the Hermitian case, for complex non-Hermitian systems in two dimensions, topological invariants can be calculated in real space. Detailed calculations can be found in Appendix \ref{sec: topological phase}. Using this method, we compute the real-space topological invariants of the loss $\gamma=[0,1]$ as a function of $g_1 (g_3)$ and $g_2 (g_4)$, resulting in the phase diagrams for gaps I and II, as illustrated in Figs. \ref{fig:two_ssh_band1}(b) and \ref{fig:two_ssh_band1}(c), respectively. Notably, gap I exhibits a nontrivial Chern number \(C = 1\), indicating the presence of a pair of topological chiral edge states. In gap II, a larger nontrivial Chern number \({C} = 2\) is observed, corresponding to two pairs of topological chiral edge states. The results for both gaps are consistent with the projected band structure. Moreover, the Chern numbers in gaps III and IV exhibit opposite signs compared to those in gaps II and I. 
\section{HYBRID SKIN-TOPOLOGICAL MODES}
\subsection{High Chern number topological edge states and hybrid skin-topological modes under $C_4$ symmetry}
\label{sec:c4 symmetry}

\begin{figure}    
\centering
\includegraphics[width=0.5\textwidth]{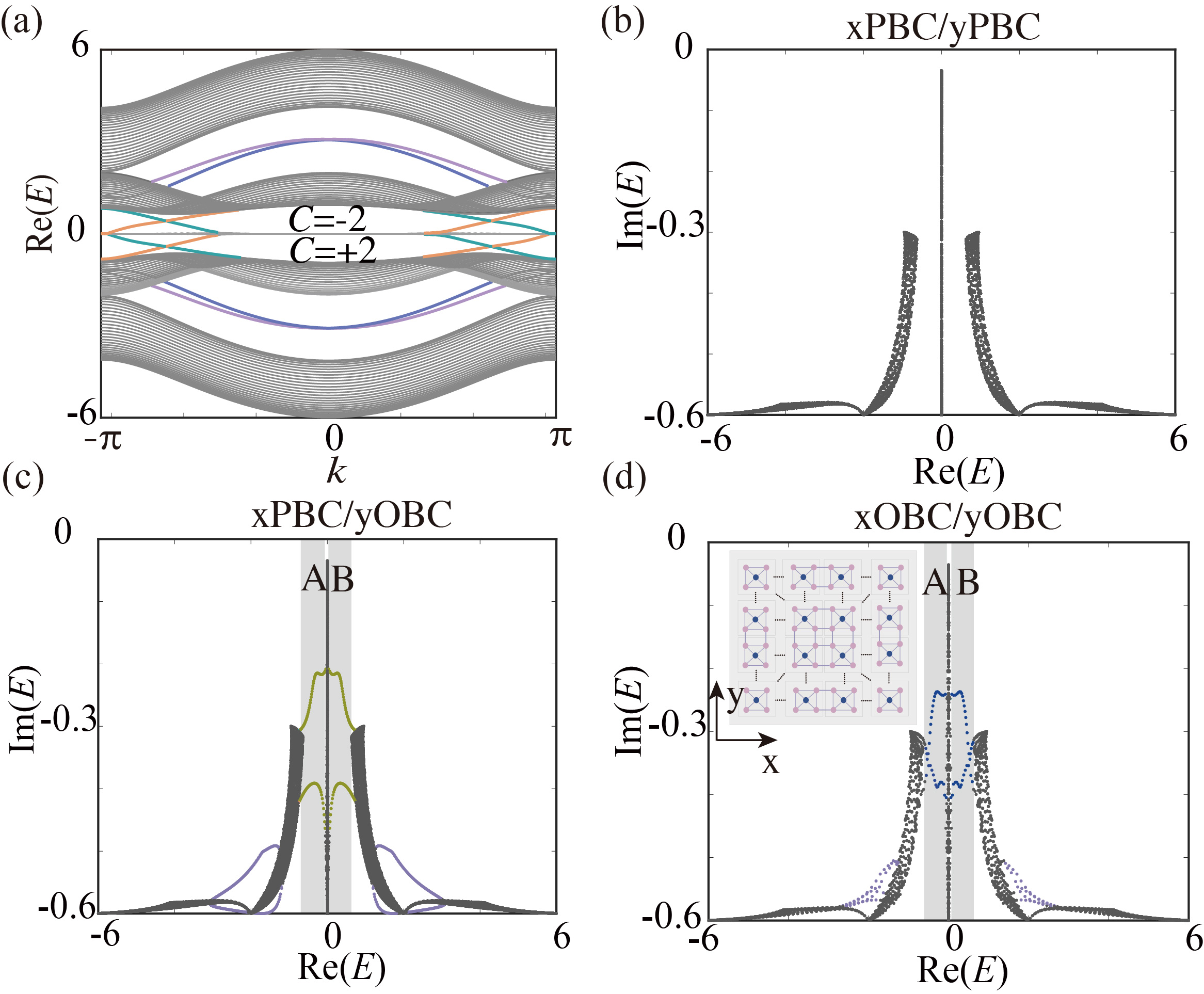}
\caption{The non-Hermitian properties of this effective model under the $C_4$ rotation symmetry. (a) The real part of the projected band structure. The Chern numbers for band gaps II and III are labeled. The edge states within band gaps II and III are shown in orange and green, respectively, depending on their localization at the upper and lower edges. (b-d) The complex energy spectra with xPBC/yPBC, xPBC/yOBC, and xOBC/yOBC, respectively. The olive dots in (c) represent  the degenerate edge states. The hybrid skin-topological corner modes in (d) are indicated in blue. The inset represents the schematic depiction of the quadrilateral supercell with xOBC/yOBC. The black dots represents omitted unit cells. The number of sites for xPBC/yOBC and xOBC/yOBC are 60 and 3380, respectively. The parameters are $t_1 = 1$, $t_2 = 2$, $g_{1, 2, 3, 4}= 0.5$, and $\gamma = -0.6$.}
\label{fig:boundy}
\end{figure}

We now beginning to investigate the non-Hermitian properties of this effective model under the $C_4$ rotation symmetry. The loss coefficient and coupling strengths are set as $\gamma = -0.6$ and $g_{1,2,3,4}=0.5$, respectively. Then,  based on the phase diagrams (Figs. \ref{fig:two_ssh_band1}(b) and (c)), one infers that two pairs of chiral edge states exist within band gaps II and III, as clearly evidenced in the real part of projected band structure depicted in Fig. \ref{fig:boundy}(a). Moreover, the projected band structure exhibits gapless, with edge states marked as dark purple and light purple in Fig. \ref{fig:boundy}(a). This situation is not further analyzed in this paper. Like in Fig. \ref{fig:two_ssh_band1}(a), we use orange (green) to represent the eigenstates localized at the upper (lower) edge. And for the same real part, $\phi_{\text{orange},1}/\phi_{\text{green},2} (\phi_{\text{orange},3}/\phi_{\text{green},4})$ is used to label the edge states when $k>0(k<0)$, which facilitates the subsequent analysis of the hybrid skin-topological modes.

The corresponding complex energy spectra for the the non-Hermitian effective model are shown in Figs. \ref{fig:boundy}(b)-\ref{fig:boundy}(d), respectively, corresponding to three different boundary conditions (xPBC/yPBC, xPBC/yOBC, and xOBC/yOBC). PBC (OBC) represents the periodic (open) boundary condition in the $x$ or $y$ direction. A schematic of the regular rectangular supercell with xOBC/yOBC is presented in the inset of Fig. \ref{fig:boundy}(d). Under different boundary conditions (as shown in Figs. \ref{fig:boundy}(b)-\ref{fig:boundy}(d)), the changes in the complex energy spectrum of the bulk continuum are minor, indicating that the first-order skin effect is almost absent in the bulk. This absence results from the cancellation of nonreciprocity arising from the symmetry of the local flux, along with a balance between gain and loss within the bulk\cite{hybrid-skin-effect1}. 

\begin{figure}[]   
        \centering
        \includegraphics[width=0.5\textwidth]{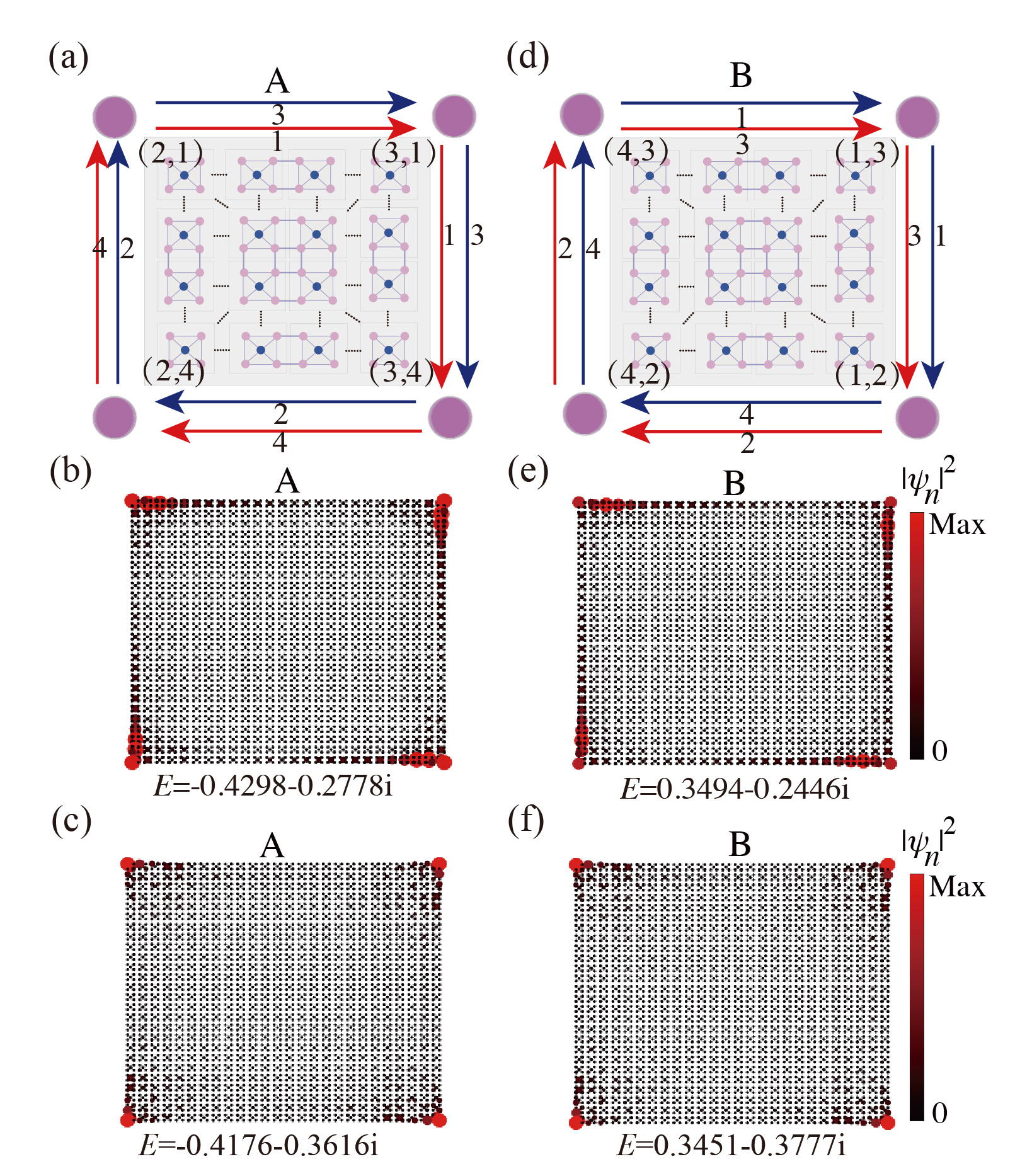}
        \caption{Non-Hermitian $C_4$ symmetric effective model with corner skin modes. (a) and (d) The schematic of corner skin modes existing in the four configurations of region A and B. Blue (red) arrows represent chiral edge states exhibiting relative attenuation (amplification), accumulating waves in their propagation (opposite) directions. Glowing purple spheres indicate potential localized corner skin modes. (b, e) and (c, f) depict the wave function distributions of representative corner skin modes with higher and lower imaginary energy in the quadrilateral supercell, corresponding to the configurations in (a) and (b). The radius of the solid circles in the supercell is proportional to the amplitude at each atomic position.}
\label{fig:skin3}
\end{figure}

Gaps II and III are labeled as regions A and B in the complex energy spectrum shown in Figs. \ref{fig:boundy}(c) and \ref{fig:boundy}(d). For the same real part, the edge states \(\phi_{\text{orange},1}(\phi_{\text{green},2})\) and \(\phi_{\text{green},4}(\phi_{\text{orange},3})\) are degenerate, marked in olive in Fig. \ref{fig:boundy}(c). The gap states are indicated in blue in Fig. \ref{fig:boundy}(d). Interestingly, for the same real energy, the imaginary energy in regions A exhibit a relationship of  $\text{Im}\text{(}E_{\mathrm{2/3}, \text{xPBC/yOBC}}\text{)}<\text{Im}\text{(}E_{\text{blue-up/-down, xOBC/yOBC}}\text{)}<\text{Im}\text{(}E_{\mathrm{1/4}, \text{xPBC/yOBC}}\text{)}$. Furthermore, all gap states are localized at the four corners, as shown in Figs. \ref{fig:skin3}(b, c, e, f). Notably, we still consider these gap states to be hybrid topological-skin modes arising from the non-Hermitian skin effect directly induced by the topological edge states. This observation is consistent with the conclusions in the literature\cite{hybrid-skin-effect4}.

For the sake of generality, this paper explains the hybrid topological-skin effect through a dynamical approach and confirms the relationship between the winding number and the NHSE based on existing analyses\cite{PhysRevLett.125.126402, PhysRevLett.124.086801, hybrid-skin-effect4}. By analyzing the interactions of edge states with dynamic behaviors in different directions at the corners, we can observe how the skin effect localizes these states at specific positions, allowing us to predict the hybrid topological-skin mode. This approach is equally applicable to the multiple pairs of edge states present in this study. The dynamics of chiral edge states depend on two key factors: the sign of the group velocity determines the propagation direction of the wave packets at the boundary; the imaginary part of the complex energy affects the lifetime of the wave packets, determining whether they are amplified or attenuated during dynamic evolution.

Taking region A as an example for detailed analysis, we define the group velocity sign as positive, with the clockwise propagation direction considered positive. The top and left sides of the rectangular supercell represent the upper edges, while the other two sides represent the lower edges. In the projected band structure shown in Fig. \ref{fig:boundy}(a), within Gap II, at the same real energy, the group velocities of the four edge states are: $+v_{\text{orange},1/3}$ and $-v_{\text{green},2/4}$. In Fig. \ref{fig:boundy}(c), within region A at the same real energy, the edge states $\phi_{\text{green},2}$ and  $\phi_{\text{orange},3}$ ($\phi_{\text{orange},1}$ and  $\phi_{\text{green},4}$)with larger (smaller) imaginary energy indicate that wave packets are relatively amplified (attenuated) in their respective directions. The propagation directions and amplification or attenuation of the four edge states along the four edges are illustrated in Fig. \ref{fig:skin3}(a). In the figures, arrows represent the propagation direction of the wave packets, with blue (red) indicating relative amplification (attenuation) along the respective directions.

In Fig. \ref{fig:skin3}(a), we observe two types of wave packets exhibiting directional amplification or attenuation along the edges. For instance, the wave packet \(\phi_{\text{orange},3}\) moving rightward along the upper edge has a longer lifetime and eventually collapses to the upper-right corner over time. Meanwhile, the downward-moving wave packet \(\phi_{\text{orange},1}\) along the right edge has a shorter lifetime, decaying exponentially. As shown by the glowing purple spheres in Fig. \ref{fig:skin3}(a), these wave packets localize at the corner, denoted as (3, 1). It is not surprising that the wave packet \(\phi_{\text{orange},1}\), moving rightward along the upper edge, has a shorter lifetime and decays more rapidly. Meanwhile, the downward-moving wave packet \(\phi_{\text{orange},3}\) along the right edge has a longer lifetime and eventually collapses to the lower-right corner after dynamic evolution. This shows that the interaction between these two wave packets at the upper-right corner does not lead to localization. Based on the above analysis, the dynamic behavior of different wave packets leads to preferences at other corners, as shown in Fig. \ref{fig:skin3}(a), corresponding to the hybrid topological-skin modes with higher and lower imaginary energy, as illustrated in Figs. \ref{fig:skin3}(b) and \ref{fig:skin3}(c), where the radii of solid circles are proportion to the amplitude strength at each atomic site. 

Similar analysis applies to region B. The corner skin mode wave functions in Figs. \ref{fig:skin3}(e) and \ref{fig:skin3}(f) correspond to the configuration shown in Fig. \ref{fig:skin3}(d). Compared to previous studies\cite{hybrid-skin-effect4}, the non-Hermitian photonic checkerboard lattices of effective model exhibits multiple hybrid topological-skin modes within a single band gap, enhancing the topological protection of the non-Hermitian system. This multi-mode topological protection mechanism holds significant application value in the development of highly robust, low-power topological devices. 

\subsection{High Chern number topological edge states and hybrid skin-topological modes under $C_2$ symmetry}
\label{sec:c2 symmetry}
\begin{figure}    
\centering
\includegraphics[width=0.5\textwidth]{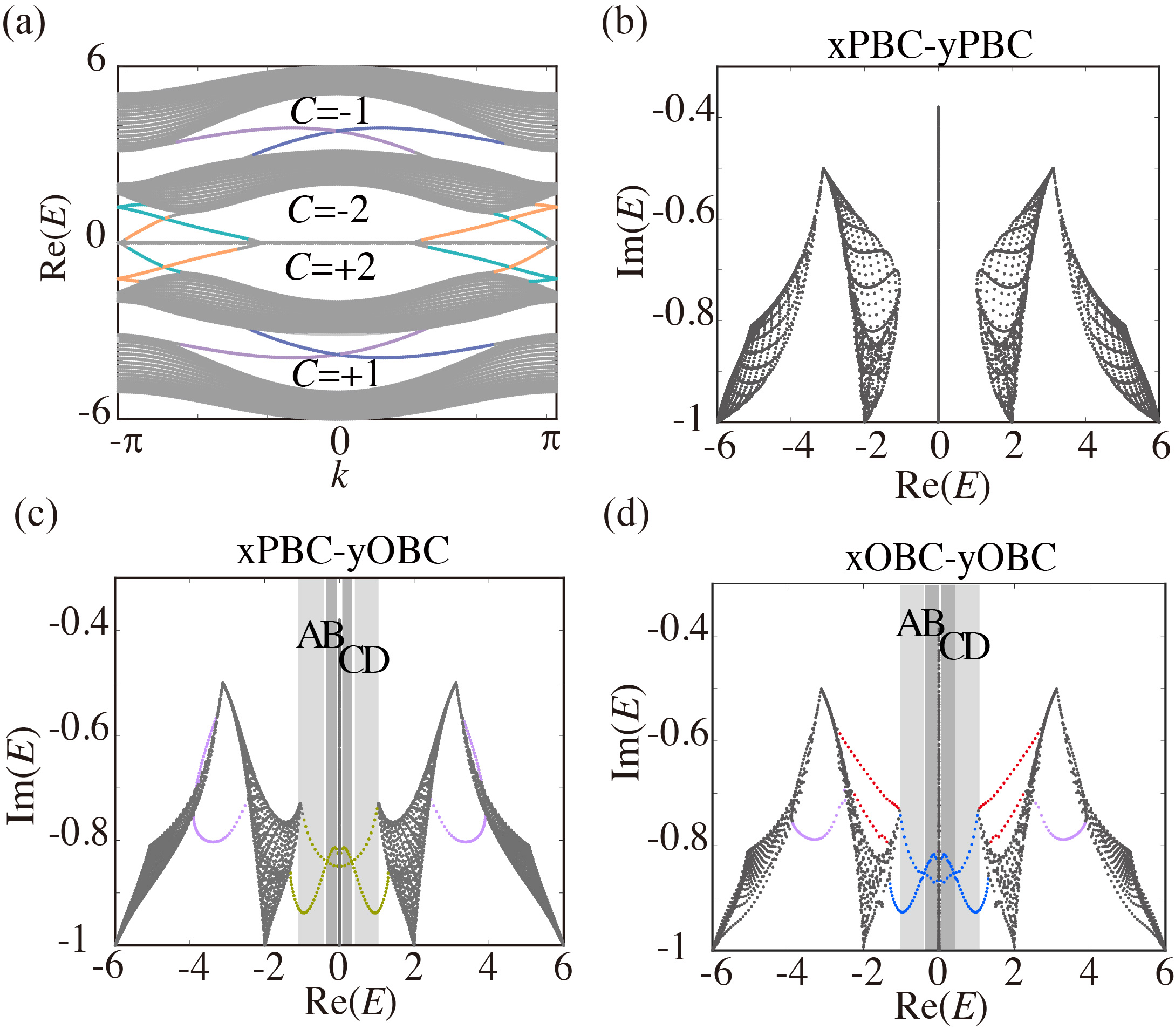}
\caption{The non-Hermitian properties of the effective model under $C_2$ symmetry. (a) The real part of the projected band structure with xPBC/yOBC. The edge states within the band gaps II and III are labeled  the same as in Fig. \ref{fig:boundy}(a). (b-d) The complex energy spectra with xPBC/yPBC, xPBC/yOBC, and xOBC/yOBC, respectively. The olive dots in (c) represent  the degenerate edge states. The hybrid skin-topological corner modes in (d) are indicated in blue. The skin-bulk modes in (d) are indicated in red. The number of sites for xPBC/yOBC and xOBC/yOBC are 60 and 3380, respectively. The parameters are $t_1 = 1$, $t_2 = 2$, $g_{1, 3}= 1$, $g_{2, 4}=2$ and $\gamma = -1$.}
\label{bobao}
\end{figure}

We will investigate the non-Hermitian effective model with $C_2$ symmetry, where the coupling strengths are $g_{1, 3}=1$ and $g_{2, 4}=2$, the dissipation coefficient is $\gamma=-1$.
As shown in Fig. \ref{bobao}(a), the real part of the projected band structure reveals that two pairs of edge states still exist in band gaps II and III, marked in the same way as before, while there is one pair of edge states in band gaps I and IV.  Figs. \ref{bobao}(b-d) correspond to the complex energy spectra under three boundary conditions: xPBC/yPBC, xPBC/yOBC, and xOBC/yOBC. Notably, the energy in both band gaps II and III exhibits crossing points, where the imaginary part of the energy reverses on either side of these crossings. Based on these crossing points, band gap II (III) is divided into regions A and B (C and D), as depicted in Figs. \ref{bobao}(c) and \ref{bobao}(d). In regions A and D, the real energy is the same, and the imaginary energy of both the upper and lower supercell states lies between the chiral edge states. In regions B and C, only the imaginary energy of the upper supercell state is positioned between the chiral edge states, while the imaginary energy of the lower supercell state is located below the edge states. This slight shift in the edge states may be attributed to the $C_{2}$ symmetry of the structure. However, it does not affect the validity of our subsequent analysis. And this clear distinction in imaginary energy distribution across different regions aids in understanding the properties of the corner skin modes that will be discussed later.

More importantly, the dynamical behavior of chiral edge states is not only influenced by the imaginary energy and the associated group velocity, but also by the weight of the chiral edge states in specific corner skin modes, denoted as $c_{j}\text{(}j=1,2,3,4\text{)}$. This factor must be considered, as in the  coupling interaction $g_{1,3}=1$ and $g_{2,4}=2$, the Chern number ${C}=2$ emerges in gaps II and III, indicating the presence of two pairs of topological edge states rather than a single pair. Moreover, the $C_{2}$ symmetry of the model alters the three factors that affect the dynamical behavior of the chiral edge states, resulting in a more diverse set of hybrid skin-topological modes. Thus, we need to redefine the previous analysis and provide a detailed derivation (see Appendix \ref{sec: wave dynamic}).

We redefine $P_j\equiv c_j\exp\text{(}E_j t_j\text{)}\text{(}j= 1, 2, 3, 4\text{)}$, used the imaginary part of the energy $E_j$, time $t_j$, and the coefficients $c_j$ to calculate the impact of wave functions during the dynamic process along different edges. In this definition, the imaginary part of the energy $E_j$ represents the difference between the edge state and the imaginary energy of the corner skin mode (as reference energy). The time $t_j$ is determined by the group velocity of the edge state, representing the time required for a wave packet to propagate a unit lattice distance along the edge. Notably, for two wave packets with longer lifetimes, their relative motion will cause them to gradually collapse toward the corner, and the product operation of $P_j$ gives the magnitude of the wave function distribution at that corner. For two wave packets with shorter lifetimes, their amplitudes will decay exponentially during their converging motion, and the product operation of the initial wave packet weight $c_j$ yields the wave function distribution at the initial corner. By comparing the wave function distributions at each corner, we can assess the manifestation of the hybrid topological skin effect at the respective corners. In addition, we also give a simple example in Appendix \ref{sec: wave dynamic} to prove the reliability of the amplifier.

\begin{figure}    
\centering
\includegraphics[width=0.5\textwidth]{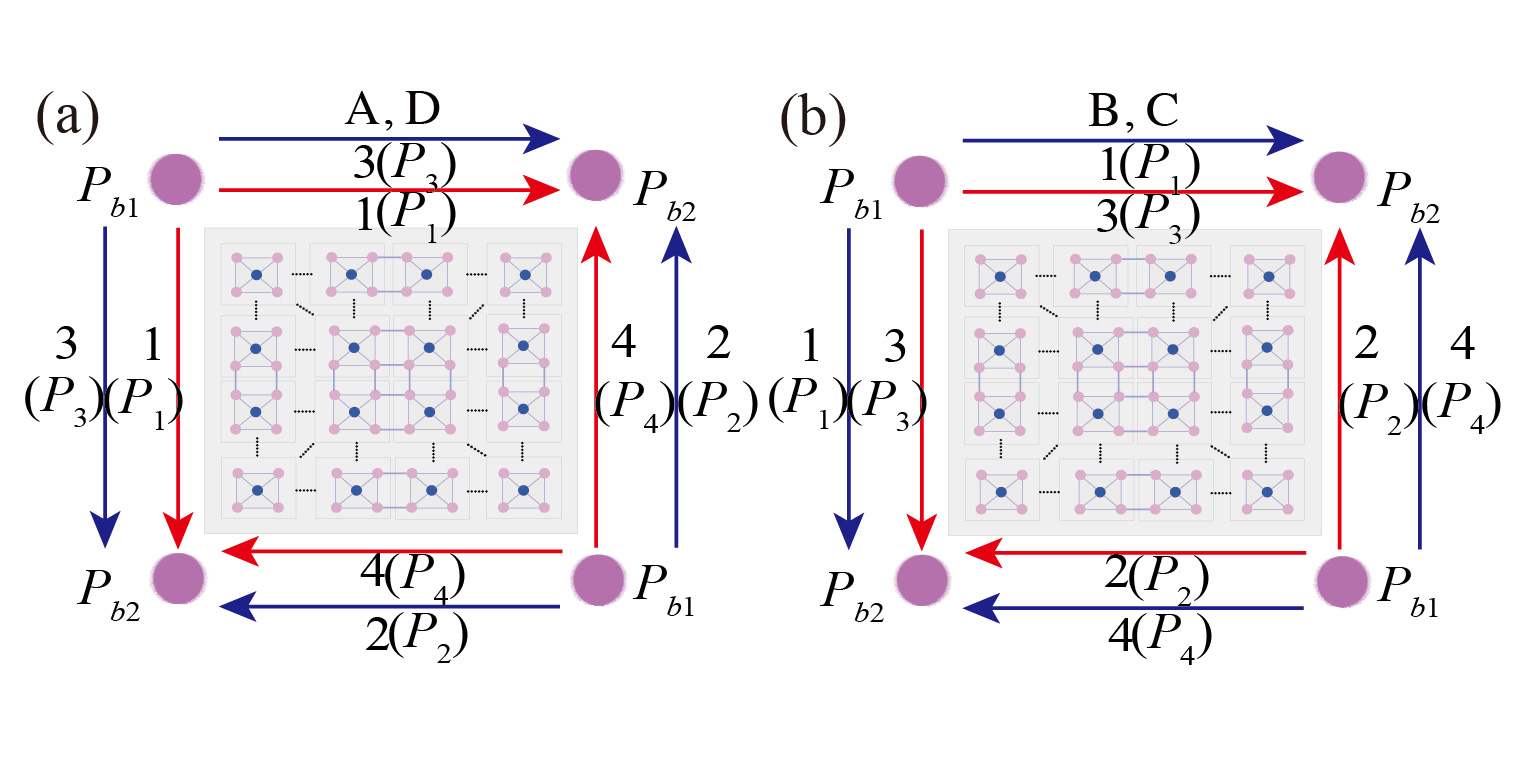}
\caption{ (a-b) Configuration diagram of the supercell showing the predicted corner skin modes in regions A, D and B, C.}
\label{fig:peizhi}
\end{figure}

\begin{figure*}    
\centering
\includegraphics[width=7.2in]{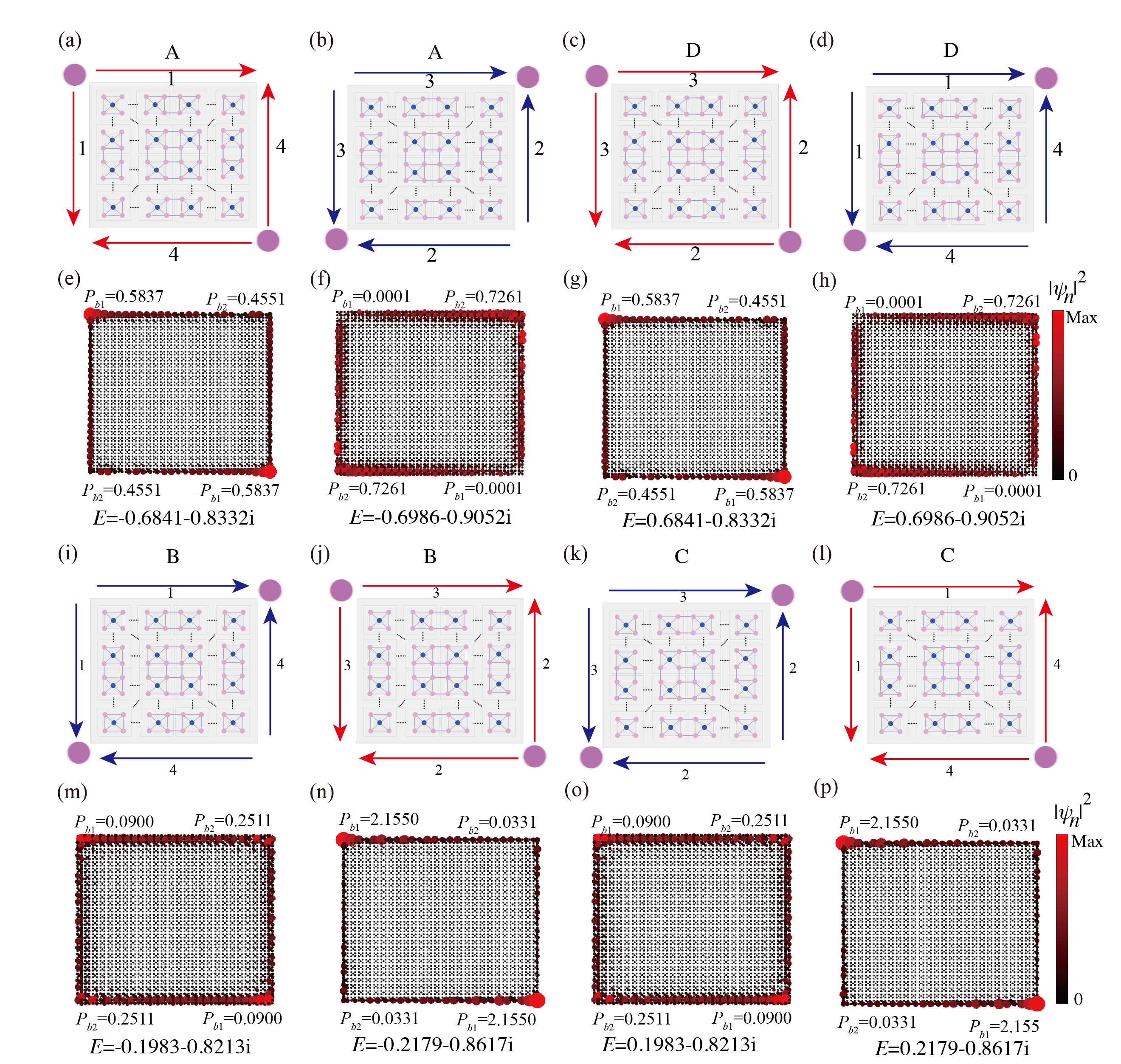}
\caption{ Non-Hermitian $C_2$ symmetric effective model with corner skin modes. (a-d) and (i-l) The schematic of corner skin modes existing in the two configurations of regions A, D and B, C. Blue (red) arrows represent chiral edge states exhibiting relative attenuation (amplification), accumulating waves in their propagation (opposite) directions. Glowing purple spheres indicate potential localized corner skin modes. (e, g, m, o) and (f, h, n, p) depict the wave function distributions of representative corner skin modes with higher and lower imaginary energy in the quadrilateral supercell, corresponding to the configurations in (a, c, i, k) and (b, d, j, l). The radius of the solid circles in the supercell is proportional to the amplitude at each atomic position.}
\label{fig:c2_skin}
\end{figure*}

We focus our analysis on region A, based on the existing analytical methods, with the relevant configuration shown in Fig. \ref{fig:peizhi}(a). For example, we select the corner skin mode with the same real energy and a higher imaginary energy ($\text{Im}\text{(}E\text{)} = -0.8332$). The two wave packets $\phi_{\text{orange},2}$ and $\phi_{\text{orange},3}$, have longer lifetimes and move along their respective edge directions, ultimately collapsing at the upper right or lower left corner after some time, resulting in $P_{b2} = P_3 \times P_2 = 0.4451$. In contrast, the two wave packets $\phi_{\text{green},1}$ and $\phi_{\text{green},4}$, have shorter lifetimes and decay exponentially, collapsing at the upper left or lower right corner at the initial moment, yielding $P_{b1} = c_1 \times c_1 (c_4 \times c_4) = 0.5837$. It is evident that $P_{b1} > P_{b2}$, thus the relevant configuration is shown in Fig. \ref{fig:c2_skin}(a). And the corner skin mode as shown in Fig. \ref{fig:c2_skin}(e). For the configurations in other regions, see Fig. \ref{fig:peizhi}.

Through a similar analysis, Figs. \ref{fig:c2_skin} (a-h) and \ref{fig:c2_skin}(i-p) were obtained, displaying the relevant configurations and corner skin effects for regions A, D, and B, C, respectively. Figs. \ref{fig:c2_skin}(a–d) and (i-l) illustrate the configuration diagrams, with each corresponding to the corner skin effects observed in Figs. \ref{fig:c2_skin}(e–h) and \ref{fig:c2_skin}(m-p), respectively. It is worth noting that, for the same real energy, the imaginary energy of the corner skin modes in Figs. \ref{fig:c2_skin}(n) and \ref{fig:c2_skin}(o) is lower than that of the edge states with xPBC/yOBC. In this case, we no longer need to reference the imaginary energy of the supercell but instead redefine $P_j$ . When the imaginary energy of the edge states with xPBC/yOBC are higher, the wave packet will grow directionally in its dynamical evolution; conversely, it will decay directionally, consistent with the first scenario discussed earlier. Similarly, we can compute \(P_{b1}\) and \(P_{b2}\) and derive the corresponding configurations, as shown in Figs. \ref{fig:c2_skin}(b, d, j, l). Here, the introduced concept \( P_j \equiv c_j \exp(E_j t_j) \) provides a more universally applicable method for analyzing multiple hybrid skin-topological modes, enriching the study of symmetry models.

\begin{figure}    
\centering
\includegraphics[width=0.48\textwidth]{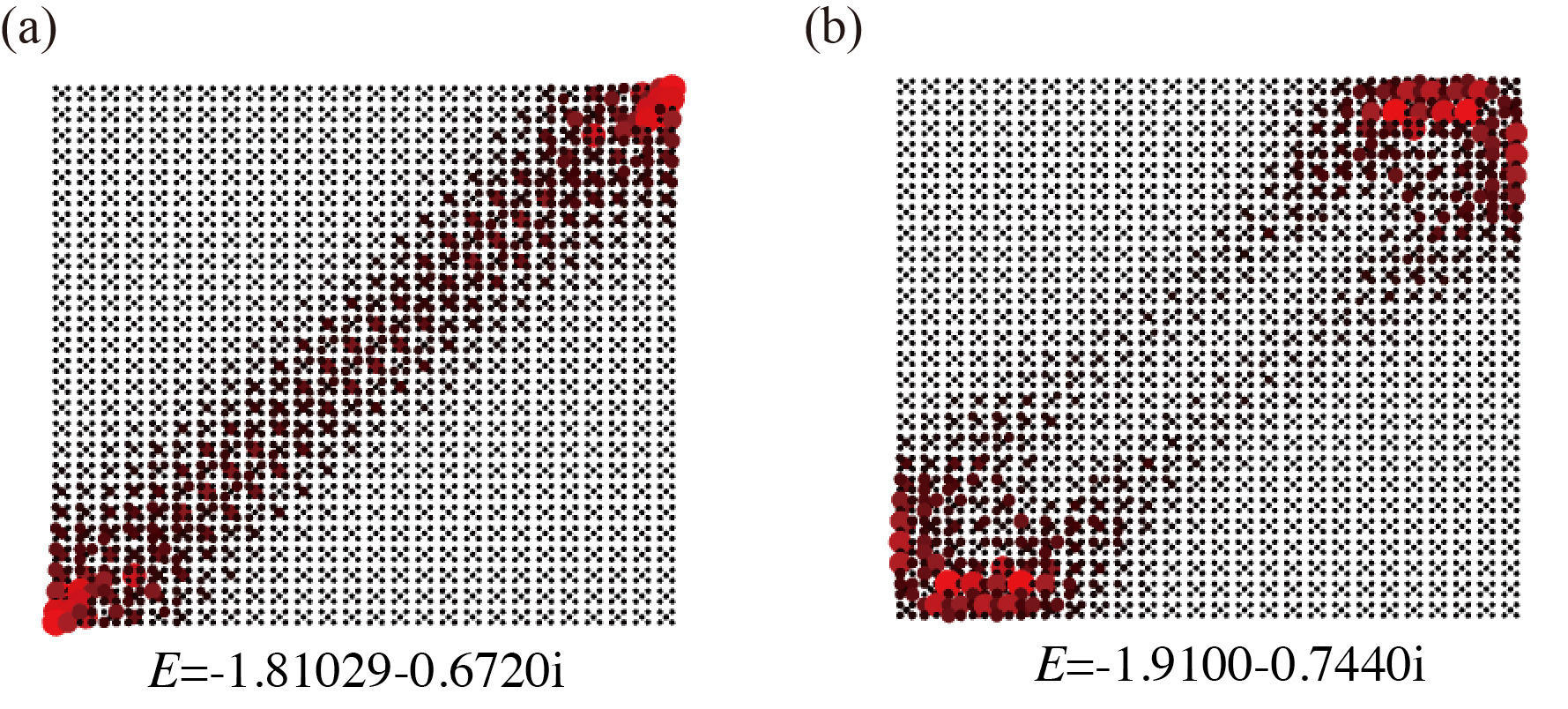}
\caption{ (a-b) The higher-order corner skin effect induced by the bulk states results in similar phenomena on both sides, where (a) selects the upper bulk state, and (b) selects the lower bulk state. }
\label{bulkstate}
\end{figure}

By comparing Figs. \ref{bobao}(c) and \ref{bobao}(d), we also observe multiple bulk states, which are marked in red. The upper bulk states with higher imaginary energy and the lower bulk states with lower imaginary energy are shown in Figs. \ref{bulkstate}(a) and \ref{bulkstate}(b), respectively, from the red regions on the left side of Fig. \ref{bobao}(d). It is evident that these corner skin modes are localized in the lower left and lower right corners, indicating that these corner skin effects are indeed caused by the bulk states. In this paper, we do not further investigate the specific localization of these corner skin modes.

\section{Conclusion and discussions}
\label{Conclusion}

We establish a non-Hermitian Chern insulator model based on checkerboard lattices. Despite its simplicity, the model encompasses rich non-Hermitian topological phases. Of special interest here, is the non-Hermitian topological band gaps with large Chern numbers which support multiple chiral edge states. In finite systems, a corner boundary can trigger the scattering among these chiral edge states without causing back scattering, making the bulk-boundary correspondence more versatile. Interestingly, we find that the corner-induced scattering leads to a higher-order non-Hermitian skin effect where the waves are trapped at the corner boundaries. Depending on the system's symmetry and parameters, the corner skin modes can appear at all the four corners or only at a pair of diagonal corners. This phenomenon is different from the previously found hybrid skin-topological effect where the system has only one branch of non-Hermitian chiral edge states and the corner skin effect is solely due to the gain and loss of the chiral edge states as determined by often by the symmetry of the system. Instead, here, we reveal through the dynamical analysis that the corner-induced scattering play an important role in the higher-order non-Hermitian skin effect, which is unique for non-Hermitian topological phases with large Chern numbers.

We emphasize that the checkerboard lattice model proposed in this work can be realized in a number of systems. First, as studied here, a photonic Floquet lattice consisting of coupled optical waveguides can realize the model, which have been realized in recent experiments based on coupled silicon photonic waveguides~\cite{Floquet-non-Hermitian4}. Second, in cold atom systems where the artificial gauge fields can be generated through the scheme of optical flux lattices~\cite{PhysRevLett.106.175301}, the checkerboard lattice model can also be realized if the dissipation is tuned~\cite{cho}.

We remark that this work not only adds to the fundamental research in non-Hermitian topological physics by enriching the mechanisms for non-Hermitian topological bulk-boundary correspondence, but also enriches the development of non-Hermitian topological photonics which may enable the realization of unconventional photonic materials and applications.

\begin{acknowledgments}
We are grateful to Dr. Weiwei Zhu valuable discussions. This work is supported by the the National Key R\&D Program of China (2022YFA1404400), the National Natural Science Foundation of China (Grant No. 12125504), and the ``Hundred Talents Program'' of the Chinese Academy of Sciences, the Gusu leading scientist program of Suzhou city, and the Priority Academic Program Development (PAPD) of Jiangsu Higher Education Institutions.
\end{acknowledgments}

\appendix
\section{ANALYZE THE EFFECTIVE HAMILTONIAN}
\label{sec: effective hamiltonian}
In Floquet systems (i.e., periodically driven systems), gauge transformations may also involve redefinition in the time domain. When analyzing the Floquet effective Hamiltonian, an appropriate gauge transformation can eliminate the time dependence, transforming it into a static equivalent description, thus simplifying the physical interpretation. We will consider artificial potential engineering as much as possible to obtain the desired coupled phases. One can consider such a Hamiltonian (\ref{Hamiltonian_total}),
 and the time-dependent Schrödinger equation is given by: 
\begin{equation}
i \partial_{z}| \psi(z)\rangle=H(z)| \psi(z)\rangle.
\label{eq:schoding}
\end{equation}
 Here, we apply a unitary gauge transformation to the evolved state $|\psi(z)\rangle=u(z)|\phi(z)\rangle$, where, $u(z)=\sum_{i,j}\text{exp}(i\Phi(z)z)\mathcal{K}_{i,j}^\dagger \mathcal{K}_{i,j}(\mathcal{K}\in\text{a, b, c, d, e})$. Plugging $|\psi(z)\rangle$ into the Eq. (\ref{eq:schoding}), we get the transformed Schrödinger equation
 \begin{equation}
i \partial_{z}| \phi(z)\rangle=H'(z)| \phi(z)\rangle,
\label{eq:schoding1}
\end{equation}
where $H'(z)=u(z)^{\dagger}H(z)u(z)+iu(z)^{\dagger}\partial_{z}u(z)$, and we can obtain
\begin{align}
H'(z)&=\sum_{i,j}t_{1}(a_{i,j}^{\dagger}b_{i,j}+d_{i,j}^{\dagger}c_{i,j}+b_{i,j}^{\dagger}c_{i,j}+a_{i,j}^{\dagger}d_{i,j})\notag\\
&+\sum_{i,j}t_{2}(a_{i+1,j}^{\dagger}b_{i,j}+d_{i+1,j}^{\dagger}c_{i,j}+b_{i,j+1}^{\dagger}c_{i,j}+a_{i,j+1}^{\dagger}d_{i,j})\notag\\
&+\sum_{i,j}[G_{1}(z)(e_{i,j}^{\dagger}a_{i,j})+G_{2}(z)(e_{i,j}^{\dagger}b_{i,j})]\notag\\
&+\sum_{i,j}[G_{3}(z)(e_{i,j}^{\dagger}c_{i,j})+G_{4}(z)(e_{i,j}^{\dagger}d_{i,j})+i\gamma(\xi_{i,j}^\dagger \xi_{i,j})]+\text{h.c.},
\label{eq:effctive2}
\end{align}
the modulation parameters: $G_m(z)=g_m\text{exp}(iA\text{sin}(\omega z+\theta_{m})),(m\in{1, 2, 3, 4})$, here, $\theta_{m}=\int_0^z(\theta_{i,j,e}-\theta_{i,j,\xi})dz'(\xi \in\text{a, b, c, d})$. 
The discrete Fourier component of Hamiltonian $H_{n}$, 
\begin{align}
H_n&=\sum_{i,j}t_{1}(a_{i,j}^{\dagger}b_{i,j}+d_{i,j}^{\dagger}c_{i,j}+b_{i,j}^{\dagger}c_{i,j}+a_{i,j}^{\dagger}d_{i,j})\notag\\
&+\sum_{i,j}t_{2}(a_{i+1,j}^{\dagger}b_{i,j}+d_{i+1,j}^{\dagger}c_{i,j}+b_{i,j+1}^{\dagger}c_{i,j}+a_{i,j+1}^{\dagger}d_{i,j})\notag\\
&+\sum_{i,j}g_{m}e^{in\theta_{m}}[(e_{i,j}^{\dagger}a_{i,j})+(e_{i,j}^{\dagger}b_{i,j})+(e_{i,j}^{\dagger}c_{i,j})+(e_{i,j}^{\dagger}d_{i,j})]\notag\\
&+\sum_{i,j}i\gamma(\xi_{i,j}^\dagger \xi_{i,j})+\text{h.c.}.
\label{eq:effctive2}
\end{align}
In the high-frequency expansion of $H_{\text{eff}}$, with
 \begin{align}
H_{\text{eff}}&=H_0+\frac{[H_{-1},H_{1}]}{\omega}+\mathcal{O}(\frac{1}{\omega^2})\notag\\
&=\sum_{i,j}t_{1}(a_{i,j}^{\dagger}b_{i,j}+d_{i,j}^{\dagger}c_{i,j}+b_{i,j}^{\dagger}c_{i,j}+a_{i,j}^{\dagger}d_{i,j})\notag\\
&+\sum_{i,j}t_{2}(a_{i+1,j}^{\dagger}b_{i,j}+d_{i+1,j}^{\dagger}c_{i,j}+b_{i,j+1}^{\dagger}c_{i,j}+a_{i,j+1}^{\dagger}d_{i,j})\notag\\
&+\frac{1}{\omega}\sum_{i,j}[g_{1}e^{i\theta_{1}}(e_{i,j}^{\dagger}a_{i,j})+g_{2}e^{i\theta_{2}}(e_{i,j}^{\dagger}b_{i,j})]\notag\\
&+\frac{1}{\omega}\sum_{i,j}[g_{3}e^{i\theta_{3}}(e_{i,j}^{\dagger}c_{i,j})+g_{4}e^{i\theta_{4}}(e_{i,j}^{\dagger}d_{i,j})]\notag\\
&+\sum_{i,j}i\gamma(\xi_{i,j}^\dagger \xi_{i,j})+\text{h.c.}
\label{eq:effctive3}
\end{align}

\section{NON-HERMITIAN TOPOLOGICAL PHASE }
\label{sec: topological phase}
So far, there are two methods to characterize the topological properties of non-Hermitian systems using topological invariants: one involves calculating the non-Bloch Chern number by integrating the Berry curvature over the generalized Brillouin zone, and the other involves computing the Chern number in real space on a finite lattice. In one-dimensional cases, the generalized Brillouin zone can be quickly determined, but for two-dimensional cases, it is generally challenging to obtain. Here, we consider the second method, which involves using the Kitaev formula to compute the Chern number\cite{hybrid-skin-effect4, PhysRevB.108.035406}. Consider a finite lattice with a circle of radius \( r \), which is divided into three regions: A, B, and C. We can get $C_{\text{real-space}}=12\pi i\sum_{j\in A}\sum_{k\in B}\sum_{l\in C}(P_{jk}P_{kl}P_{lj}-P_{jl}P_{lk}P_{kj})$. Here, \( j \), \( k \), and \( l \) represent the positions of these three regions, respectively. And $\hat{P}=\sum_{E_n<E_f}| \psi(n,R)\rangle\langle\psi(n,L)|$is the projection operator, $E_n$ is the $n$-th eigenenergy and the $E_{f}$ is Fermi energy.  Finally, the real-space Chern numbers of the model can be calculated (see Figs. \ref{fig:two_ssh_band1}(b-c)).

\section{ WAVE PACKETS DURING DYNAMIC EVOLUTION OF THE EIGENSTATE  }
\label{sec: wave dynamic}
Here,The final state $| \psi(t,n)\rangle$ can be obtained from any given moment as:
\begin{equation}
\begin{aligned}
| \psi(t,n)\rangle=U(t)| \psi(t_{0},n)\rangle.
     \label{Ham_obc}
     \end{aligned}
\end{equation}
where, $U(t)\equiv e^{-i H_{OBC}t}$ represents the time evolution operator, By exciting an  initial state $\psi(t_{0},n)\rangle$ using a Hamiltonian $H_{\text{OBC}}$ under OBC, the final state $| \psi(t,n)\rangle$ is obtained through time evolution. Next, we can perform a Fourier transform of the real-space state $| \psi(t,n)\rangle$ into the general \(k\)-space $| \psi(t,k)\rangle$,
\begin{equation}
\begin{aligned}
| \psi(t,k)\rangle=\sum_{n=1}^{N}| \psi(t,n)\rangle e^{-ikx_{n}}.
     \label{Hamiltonian_symmetry}
     \end{aligned}
\end{equation}
Thus, the final state $| \psi(t,k)\rangle $with contributing wave vectors $k$during the evolution process can be decomposed into the eigenstates of the non-Bloch Hamiltonian, and the component coefficients $c_{j}(t,k)$  can be calculated,
\begin{equation}
\begin{aligned}
| \psi(t,k)\rangle&=\sum_{j}c_{j}(t,k)| \phi_{j,R}(k)\rangle,\\
c_{j}(t,k)&=\langle \phi_{j,L}(k)|\psi(t,k)\rangle,
     \label{cxishu}
     \end{aligned}
\end{equation}
where $j$ represents the degrees of freedom of the non-Bloch Hamiltonian, and  the right and left eigenvectors of the non-Bloch Hamiltonian are $| \phi_{j,R}(k)\rangle$and $\langle \phi_{j,L}(k)|$, respectively, and they satisfy the biorthogonality condition
\begin{equation}
\begin{aligned}
\langle \phi_{i,L}(k)|\phi_{j,R}(k)\rangle=\delta_{i,j}.
     \label{Hamiltonian_symmetry}
     \end{aligned}
\end{equation}
Therefore, at any moment, Eq. (\ref{Ham_obc}) can be rewritten by analyzing the contributions to the eigenstates of the Hamiltonian $H_{\text{OBC}}$ quantities for the final state,
\begin{equation}
\begin{aligned}
| \psi(t,k)\rangle&=\sum_{j}\langle \phi_{j,L}(k)|\psi(t_{0},k)\rangle \mathrm{exp}(-iE_{j}t)| \phi_{j,R}(k)\rangle \\
&=\sum_{j}c_{j}(t,k)e^{-i\text{Re}(E_{j})t}e^{\text{Im}(E_{j})t}| \phi_{j,R}(k)\rangle.
     \label{Ham_obc1}
     \end{aligned}
\end{equation}
At the this point, $E_{j}=-i\text{Re}(E_{j})+\text{Im}(E_{j})$ is a complex energy. When considering the magnitude of the imaginary energy, we can define $P\equiv c_{j}\text{exp}(\text{Im}E_{j}t)$,which means that the magnitude of the $P$ value determines the directional amplification or decay effect during the dynamic evolution of the eigenstates.

\begin{figure}    
\centering
\includegraphics[width=0.50\textwidth]{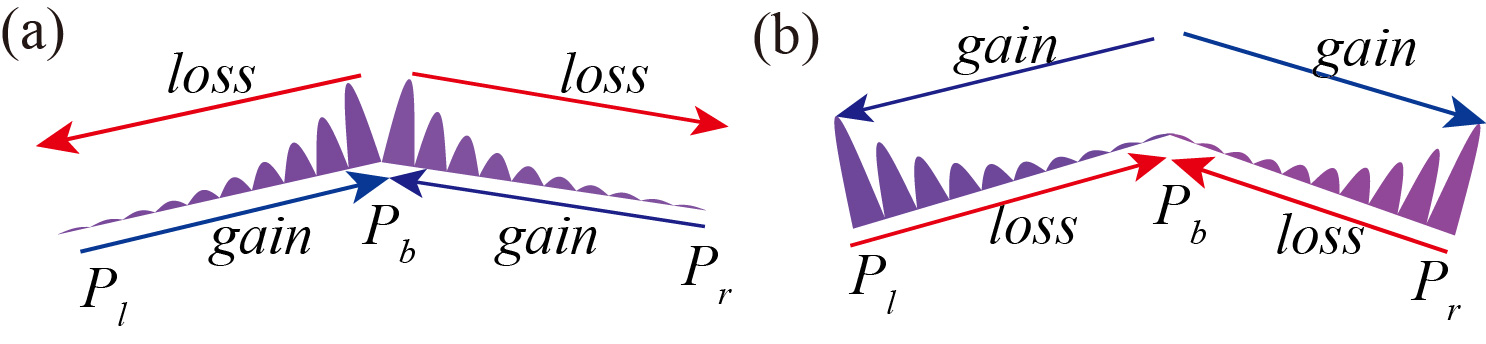}
\caption{ (a-b) Schematic diagram of whether the wave packet exhibits localization phenomena, such as directional amplification or attenuation, during dynamic evolution.}
\label{fig:gain_loss}
\end{figure}

Subsequently, we will consider two edges intersecting at a midpoint and investigate the corner skin modes at this intersection (Fig.\ref{fig:gain_loss}). Without loss of generality, we assume that the number of unit cells along the two edges is equal. First, we consider the relative motion of the wave functions of two wave packets along the profile. Suppose one wave packet starts from position$P_l$ with a group velocity $v_1$, and the other starts from position $P_r$ with a group velocity $v_2$. Here, $v_1$ and $v_2$ can be determined by calculating the dispersion of the edge states in the projected band structure. The time $t_1$ and $t_2 $ taken for the relative motion of the two wave packets can be calculated by dividing the lattice constant by the group velocity of the targeted projected band. The target imaginary energy of the edge states under investigation can be denoted as $E_+\text{(}E_-\text{)}$, representing the positive (negative) value of the energy. More importantly, the contributions $c_j$ of these two wave packets to the target edge states can be obtained using formula (\ref{cxishu}). 

Thus, we can further analyze the scenario when the imaginary component of the edge energy is $E_+$. If $P_l = c_1\text{exp}\text{(}E_+ t_1\text{)}>c_1$ and $P_r = c_2 \text{exp}\text{(}E_+ t_2\text{)} > c_2$, then $P_b = \text{exp}\text{(}E_+ t_1\text{)} \text{exp}\text{(}E_- t_2\text{)}>P_l(P_r)$, indicating that the wave function profile reaches its maximum $P_b$ at the midpoint, showing a wave packet accumulation. Conversely, when the imaginary component of the edge energy is $E_-$, we calculate \( P_l = c_1 \mathrm{exp}(E_- t_1) < c_1 \) and \( P_r = c_2 \mathrm{exp}(E_- t_2) < c_2 \), resulting in \( P_b = \mathrm{exp}(E_+ t_1) \mathrm{exp}(E- t_2) < P_l(P_r) \), showing no wave packet accumulation at the midpoint.

 We consider the case where the wave functions of two wave packets move toward each other along the profile. Suppose two wave packets start from position \( p_l \), with one moving along the left edge with a group velocity \( v_1 \), and the other along the right edge with a group velocity \( v_2 \). Based on the above analysis, when the imaginary component of the edge energy is \( E_- \), assuming \( P_b = c_1(c_2) \), \( P_l = c_1 \mathrm{exp}(E_- t_1) < c_1 \), and \( P_r = c_2 \mathrm{exp}(E_-t_2) < c_2 \), it is evident that \( P_b > P_l(p_r) \), indicating a maximum wave function profile at the midpoint. Conversely, if the imaginary component of the edge energy is \( E_+ \), then \( P_l = c_1 \mathrm{exp}(E_+ t_1) > c_1 \), \( P_r = c_2 \mathrm{exp}(E_+ t_2) > c_2 \), indicating that \( P_b < P_l(P_r) \), and thus, no accumulation occurs at the midpoint.

Based on the above analysis, combined with our actual quadrilateral model where each corner represents a $ P_{bi} $value, the calculations ultimately show that when the \( P_{bi} \) value is highest, a collective skin mode will appear at the designated corner. On the other hand, when the \( P_{bi} \) values at other corners are lower, the collective skin mode will either not appear or be relatively less pronounced.

When calculating \(P_{bj}\) in the actual model, two parameters, \(E_j\) and \(t_j\), must be determined. We will provide further explanation below. When considering the same real part of the energy \(\text{Re}(E_j)\), we define \(\delta E = \text{Im}(E_{j,\text{xPBC/yOBC}}) - \text{Im}(E_{\text{supercell}})\), where the imaginary energy of the supercell \(\text{Im}(E_{\text{supercell}})\) is taken as the reference, and \(E_{j,\text{xPBC/yOBC}}\) represents the imaginary energy of the different edge states \(j\) under xPBC/yOBC. If \(\Delta E > 0\), the wave packet located at the boundary will continuously amplify during its dynamic evolution. Conversely, if \(\Delta E < 0\), the wave packet at the boundary will decay over time.

And to explain the time \(t_j\), for each \(E_{j,\text{xPBC/yOBC}}\), there is a corresponding wave vector \(k_j\) on the projected band structure, and the group velocity can be calculated as \(v_j = \partial Re[E_j(k)]\). Assuming the number of unit cells on each boundary is consistent to simplify the calculation, the time is defined as \(t_j = 1/v_j\).

\nocite{*}

\bibliography{NH_chiral_topological}

\end{document}